\documentclass[prx, %
 reprint,
 amsmath,amssymb,
 aps,
]{revtex4-2}

\usepackage{graphicx}
\usepackage{dcolumn}
\usepackage{amsmath}
\usepackage{xcolor}
\usepackage{multirow}
\usepackage{soul}

\newcommand{\sco}{$\alpha$-SrCr$_2$O$_4$}
\newcommand{\cm}{cm$^{-1}$}

\begin{document}

\preprint{APS/123-QED}

\title{Tuning J$_1$-J$_2$ in Quasi-2D Triangular Lattice Antiferromagnet $\alpha$-SrCr$_2$O$_4$ via Uniaxial Pressure}

\author{Jazzmin Victorin}
\author{Shreenanda Ghosh}
\author{Seyed Koohpayeh}
\author{Chris Lygouras}
\author{Natalia Drichko}
\email{Corresponding author: Drichko@jhu.edu}
\affiliation{Department of Physics and Astronomy, The Johns Hopkins University, Baltimore, MD, 21218, USA}

\date{\today}

\begin{abstract}

Triangular lattice antiferromagnets first attracted attention as a frustrated magnetic lattice which can serve as a platform to realize the resonating valence bond state. While the triangular lattice itself was shown to support classical 120 degree order, many theoretical phase diagrams suggest a quantum spin liquid state within a small range of parameters. One possible avenue to achieve such a state is to tune the anisotropy of the triangular lattice antiferromagnet by applying uniaxial pressure. This motivated our Raman scattering study of quasi-two-dimensional antiferromagnet $\alpha$-SrCr$_2$O$_4$ under applied uniaxial pressure. Under ambient conditions, $\alpha$-SrCr$_2$O$_4$ develops long-range helical magnetic order below T$_N$ = 43 K. We identify two-magnon excitations associated with this long-range antiferromagnetic order below T$_N$ at 15.5 meV and 40 meV by comparison with spin wave calculations. We observe the two features from the two-magnon excitation shift away from (towards) each other under applied tensile (compressive) pressure, indicating a decrease (increase) in anisotropy. Raman active phonons show a shift to higher (lower) frequencies under applied compressive (tensile) pressure, indicating efficient transmission of pressure and tuning of the lattice. We show spin wave and two-magnon density of states calculations under uniaxial pressure are consistent with our experimental results. 

\end{abstract}

\maketitle

\section{\label{intro}Introduction}

The triangular lattice antiferromagnet is the simplest example of a system emblematic of magnetic frustration \cite{Balents2010}. While the first state to motivate studies of this lattice, the valence bond solid \cite{Anderson1973,Ueda1999}, has not been fully realized, triangular lattice antiferromagnets still demonstrate a rich range of magnetic phases. An isotropic triangular lattice Heisenberg antiferromagnet (TLHAF) orders with spins oriented at 120$^\circ$, but such order can be destabilized by next nearest neighbor interactions \cite{Diep1994}, ring exchange \cite{Motrunich2005}, or magnetic anisotropy due to spin-orbit coupling \cite{Chernyshev2019,Balents2018}. Some of the phases resulting from these extended interactions are predicted to show a spin liquid state \cite{Sorella2006}, with experimental confirmation actively discussed in the community \cite{Saito2003,Scheie2024}. Many theoretical calculations suggest phase diagrams that explore the magnetic ground state of the TLHAF as a function of an applied external magnetic field \cite{Shannon2011,Balents2013,Imada2014}; however, in terms of accessibility and controllability, it would be more advantageous to continuously tune magnetic properties by mechanically distorting the lattice to tune in-plane anisotropy. The phase diagram of the TLHAF, where exchange anisotropy J$_2$/J$_1$ is a tuning parameter, was suggested to lead to a quantum disordered state \cite{Merino1999}, and more recent theoretical studies suggest that tuning J$_2$/J$_1$ could result in a tuning of the ground state on an isotropic lattice from 120$^\circ$ order, to helical order, and even to stripe order \cite{Chernyshev2025}.  

The theoretical studies considering different ground states of triangular lattice antiferromagnets as a function of anisotropy, as briefly discussed above, do not take into consideration factors important for determining the ground state such as magneto-elastic coupling, which can lead to unconventional freezing and glassy behavior in triangular lattice antiferromagnets \cite{Nakatsuji2005,Liebman2024}. Such magneto-elastic coupling can be strongly affected by a change of anisotropy, potentially affecting the magnetic ground state. In the last decade, applying uniaxial pressure has become a popular and proven method to mechanically tune fundamental properties in strongly correlated systems \cite{Chu2021, Xu2022, Klauss2021}. Additionally, application of uniaxial pressure allows us to experimentally address magneto-elastic coupling and its role in the stabilization of the magnetic ground state.  

In this work, we tune the anisotropy of spin S=3/2 triangular lattice Heisenberg antiferromagnet \sco ~ via uniaxial pressure. While \sco ~is a weakly anisotropic TLHAF, the frustration in this material originating from slightly distorted triangular lattice layers of S=3/2 Cr$^{3+}$ ions makes the overall magnetism in this compound quite intriguing. The Curie-Weiss constant $\Theta_{CW} \approx$ -596 K for \sco ~\cite{Cava2011}, but the material develops long-range incommensurate helical magnetic order with spins confined to the \textit{ac} plane below T$_N \approx$ $43~K$ \cite{Valentine2015,Cava2011,Damay2017} indicating a strong suppression of ordering as a result of frustration $f = |\Theta_{CW}|/T_C \approx 14$. 

\begin{figure}
    \centering
    \includegraphics[width=\linewidth]{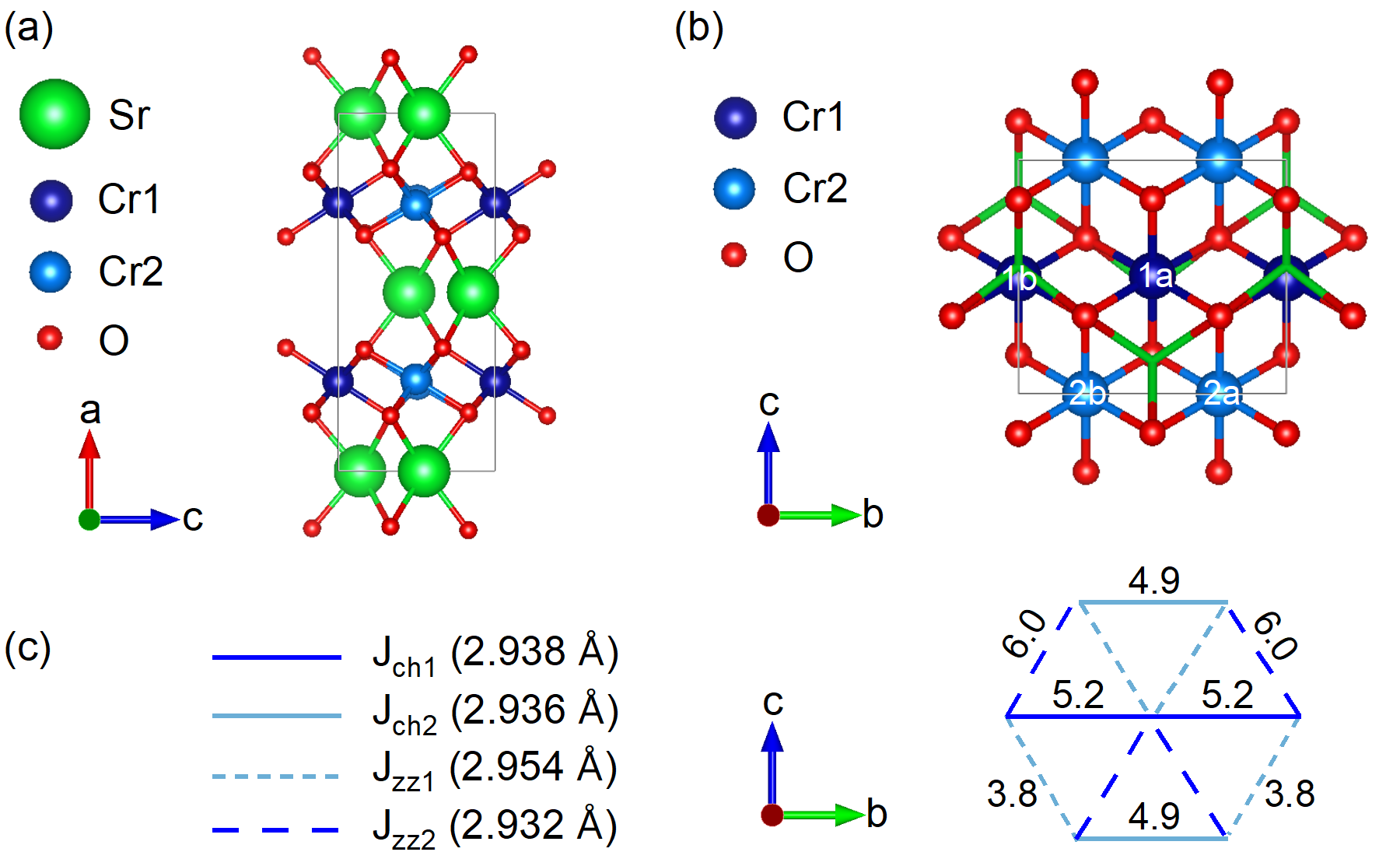} 
    \caption{(a) Crystal structure of quasi-2D TLHAF $\alpha$-SrCr$_2$O$_4$ out-of-plane. Light and dark blue spheres denote two in-equivalent Cr$^{3+}$ sites. (b) The \textit{bc} plane of $\alpha$-SrCr$_2$O$_4$ with Sr atoms omitted for clarity. (c) Simplified schematic view of exchange interactions on triangular lattice of Cr$^{3+}$ ions. Values in schematic in units of meV.}
    \label{fig:structure}
\end{figure}

Low dimensional physics in conjunction with multiferroic behavior \cite{Wu2012} make this an attractive material to probe under applied uniaxial pressure {\color{blue}as strongly coupled magnetic and lattice degrees of freedom can lead to exotic mixed magnon-phonon excitations such as the electromagnon, which is an excitation that has the character of a magnon with electric dipole coupling to electromagnetic radiation \cite{Drew2007}.} We use Raman scattering spectroscopy to follow the change of magnetic excitations and phonons, including those involved in magneto-elastic coupling \cite{Valentine2015}. Raman spectroscopy is a valuable tool in studies of frustrated magnetism under uniaxial pressure due to its sensitivity to local lattice distortions and magnetic exchange interactions through optical phonons and magnons \cite{Hackl2007}, respectively, as well as its unique sensitivity to magneto-elastic coupling \cite{Tsurkan2009}. We demonstrate a tuning of the $J_2/J_1$ ratio, which in this material we define as $J_{ch2}/J_{zz2}$ (see Fig. \ref{fig:structure} (c)), away from 0.82 observed at ambient pressure. While observed changes in magnetic excitations are small, our spin wave calculations can put the limits on the change of magnetic anisotropy we observe under pressure applied in-plane along the \textit{b} axis.  

\section{\label{methods} Experimental Methods} 

\subsection{Sample and Strain Configuration} 

Single crystals of \sco\ were grown using an optical floating zone method detailed in \cite{Valentine2015}. Quasi-two-dimensional (2D) material \sco ~is a S=3/2 weakly anisotropic triangular lattice Heisenberg antiferromagnet (see Fig. \ref{fig:structure}). The crystal has an orthorhombic structure due to two in-equivalent Cr$^{3+}$ sites that form distorted triangular layers of S=3/2 Cr$^{3+}$ in the \textit{bc} plane (see Fig. \ref{fig:structure}(b) and \ref{fig:structure}(c)). Due to crystallographic twinning in the \textit{bc} plane \cite{Valentine2015,Damay2017}, it was not possible to distinguish the \textit{b} axis from the \textit{c} axis during crystal alignment using X-ray Laue or Raman Scattering Spectroscopy. Even so, the direction of applied uniaxial pressure was estimated by the response of the two-magnon excitation. The direction of applied uniaxial pressure was determined to be along the \textit{b} axis from the experiment and comparison to theoretical spinwave calculations. More details are provided in Section \ref{theory}. 

\begin{figure}
    \centering
    \includegraphics[width=\linewidth]{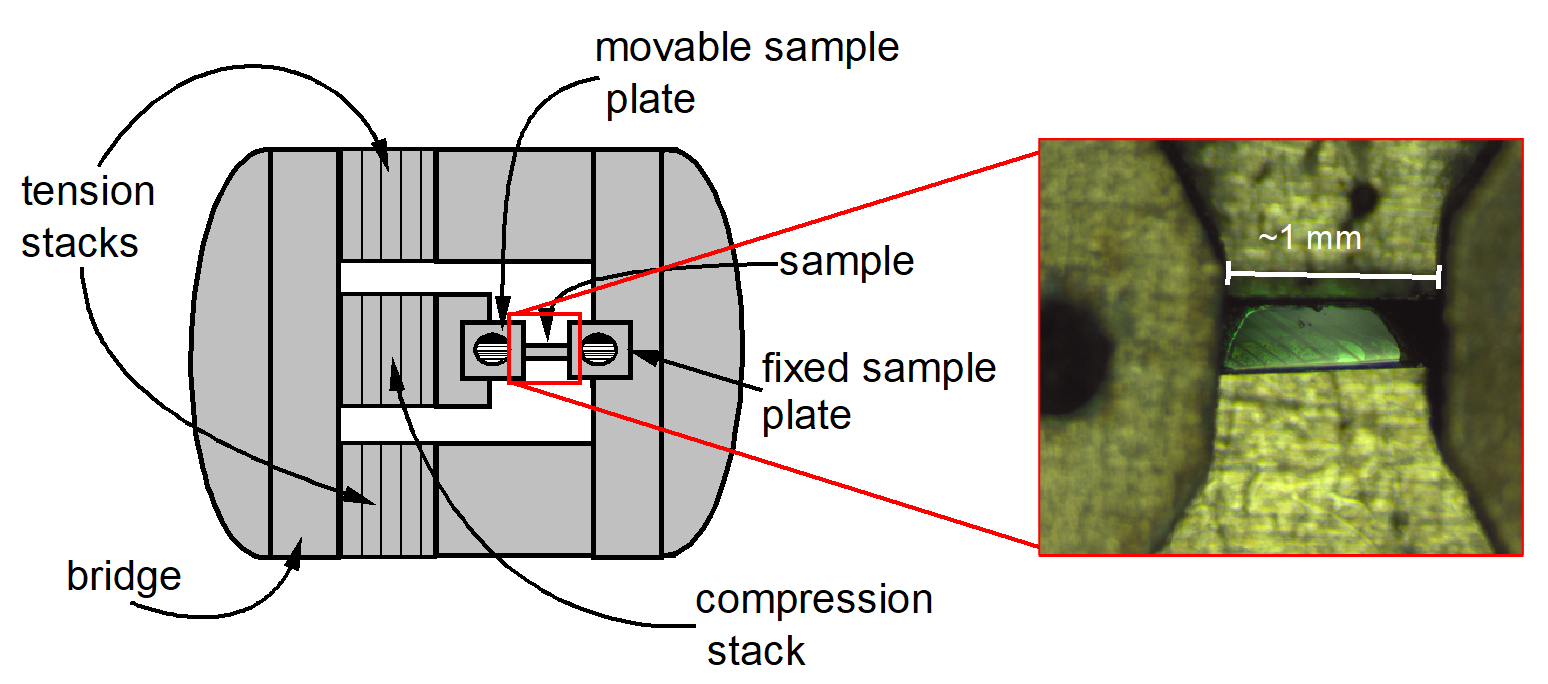} 
    \caption{Schematic of CS100 Razorbill strain cell and optical microscope image of bar-shaped $\alpha$-SrCr$_2$O$_4$ sample used for uniaxial pressure measurements.}
    \label{fig:cellandsample}
\end{figure}

A 1 mm x 0.4 mm x 0.2 mm rectangular piece of \sco ~was cut along the $<$010$>$/$<$001$>$ direction with a wire saw and sandwiched with stycast between two titanium sample plates in a floating configuration (see Fig. \ref{fig:cellandsample}) and mounted in the CS100 piezoelectric cell from Razorbill Instruments, which allows us to apply tensile and compressive uniaxial pressure in-situ and continuously tune properties at low temperatures \cite{Mackenzie2014}. In order to do measurements in the temperature range from 300 to 20 K, the CS100 strain cell was fixed onto the cold finger of a Janis ST-500 cryostat. Measured capacitance readings from the Razorbill piezoelectric cell were converted into displacement using the following equation: 

\begin{equation}
    C = \frac{\varepsilon_0 A}{d + d_0} + C_p
\end{equation}

where d is displacement, d$_0$ is the displacement at zero voltage, and C$_p$ is a parallel capacitance term to account for imperfections in the cell. This calculated displacement was then used to estimate pressure applied to the sample using 

\begin{equation} 
    \varepsilon = \frac{\Delta L}{L_{eff}} 
\end{equation} 

where L$_{eff}$ is the effective length of the sample. Strain transmission was also determined using the response of the high frequency phonons under applied uniaxial pressure. This method is detailed in Section \ref{lattice}.

\subsection{Raman Spectroscopy Measurements} 

Raman scattering spectroscopic measurements were carried out using the Horiba Jobin-Yvon T64000 triple monochromator spectrometer equipped with a liquid nitrogen cooled CCD detector. Raman scattering was excited with the 520.8 nm excitation line of an Ar$^+$-Kr$^+$ laser in the pseudo-Brewster’s angle geometry with $10.5~mW$ of laser power and elliptical laser probe of about $50\times100~\mu$m. Spectra were recorded in the frequency range from 55 to 690 \cm ~with spectral resolution 2 \cm. All spectra were normalized by 1 + $n(\omega,T)$, where $n(\omega,T)$ is the Bose-Einstein thermal factor, $n(\omega,T)$ = [exp$(\hbar\omega/k_BT) - 1]^{-1}$. 

\section{\label{lattice}Lattice Response to Temperature and Uniaxial Pressure}

Higher frequency oxygen phonons of $\alpha$-SrCr$_2$O$_4$ are not coupled to the magnetic excitation \cite{Valentine2015}, and thus can be used to follow the change of the lattice as a function of temperature and uniaxial pressure. It is well known that phonon frequency depends on the lattice parameter as 

\begin{equation}
    \omega(T) = \omega_0 + \Delta\omega^{(1)}(T) + \Delta\omega^{(2)}(T)
    \label{eq:freq}
\end{equation}

where $\omega_0$ is the bare harmonic frequency, $\Delta\omega^{(1)}(T)$ is the anharmonic thermal expansion term, and $\Delta\omega^{(2)}(T)$ is the anharmonic phonon-phonon coupling term. The thermal expansion contribution $\Delta\omega^{(1)}(T)$ is given by 

\begin{equation}
    \Delta\omega^{(1)}(T) = \omega_0 [\text{exp}(-\gamma \int[\alpha_a(T') + \alpha_b(T') + \alpha_c(T')]dT') - 1]
\end{equation}

where $\gamma$ is the Gr\"uneisen parameter and $\alpha_{a,b,c}(T)$ is the coefficient of linear thermal expansion along the a, b, and c axes, respectively: $\alpha_{L}(T)$ = $\frac{1}{L} \frac{dL}{dT}$. The phonon-phonon coupling contribution is given by 

\begin{equation}
    \Delta\omega^{(2)}(T) = B(1+2n(\omega,T))
\end{equation}

where B is a fitting parameter, $\omega$ = $\omega_0/2$, and $n(\omega,T)$ is the Bose-Einstein thermal factor \cite{Smirnov2012}. 

We observe hardening of phonons in the Raman spectra of \sco ~obtained on cooling under ambient pressure \cite{Valentine2015}, which is indicative of lattice contraction and an increase of lifetime. This behavior follows Eq. \ref{eq:freq} and allows us to extract the Gr\"uneisen parameter for these phonons, using the known contraction of the unit cell on cooling \cite{Cava2011}. Using the extracted Gr\"uneisen parameter, the strain gauge factors for these phonons were calculated using  

\begin{equation}
    \gamma = \frac{|\partial \omega/\partial\varepsilon|}{\omega_0(1-\nu)}
    \label{eq:gaugefactor}
\end{equation}

where $\nu$ is the Poisson ratio \cite{Ferrari2009}, which was assumed to be 0.2 in this compound \cite{Son2016}. 

Using the temperature dependence of the A$_g$ 602 \cm\ phonon's frequency at ambient pressure outside of the strain cell \cite{Valentine2015}, we extracted the Gr\"uneisen parameter $\gamma$ = 0.45 $\pm$ 0.18, and the resulting gauge factor was determined to be 2.19~\cm/\%. 

\begin{figure}
    \centering
    \includegraphics[width=\linewidth]{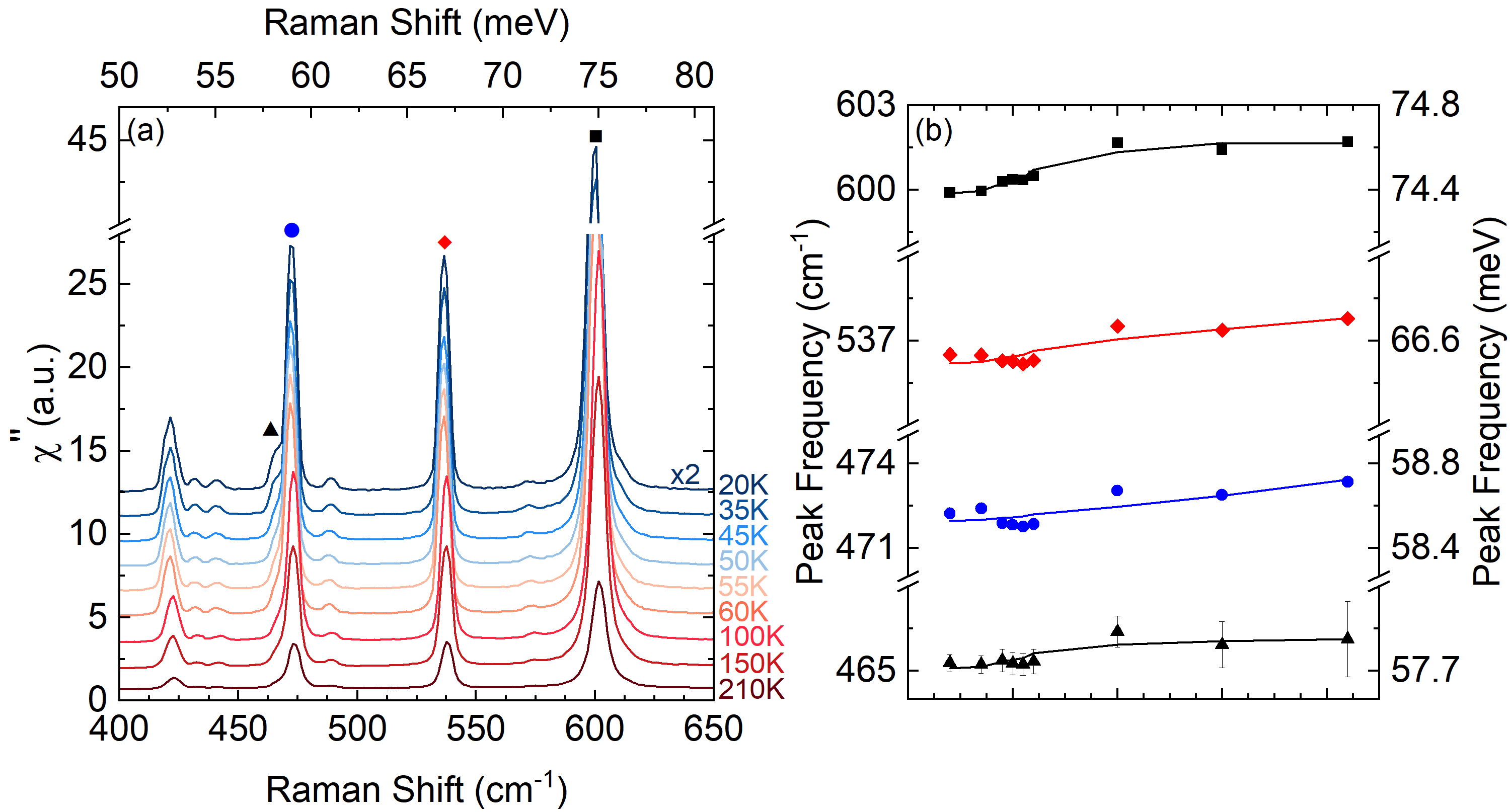} 
    \caption{(a) Temperature dependence of \sco ~sample mounted in a strain cell in the spectral range of oxygen phonons between 400 and 650 cm$^{-1}$. Unpolarized Raman spectra shifted along the y axis for clarity. (b) Frequencies of select oxygen-related phonons marked in panel (a). Solid lines are fits to symmetrical three-phonon coupling model (Eq. \ref{eq:freq}).  
    \label{fig:tdep_high_freq}}
\end{figure}

In Fig. \ref{fig:tdep_high_freq} we present spectra in the range between 400 and 650 \cm ~in the temperature range between 300 and 20 K obtained for the sample fixed in the strain cell with no voltage applied to the piezos. We observe phonons shift to lower frequencies upon cooling in the strain cell in contrast to the phonon hardening observed in the sample outside of the strain cell indicating the sample in the strain cell is undergoing tensile stress as a result of the differential thermal expansion between the sample and titanium of the strain cell environment. This tensile stress is estimated to be $\sim$0.41\% using the approximated gauge factor above. 

\begin{figure} 
    \centering
    \includegraphics[width=\linewidth]{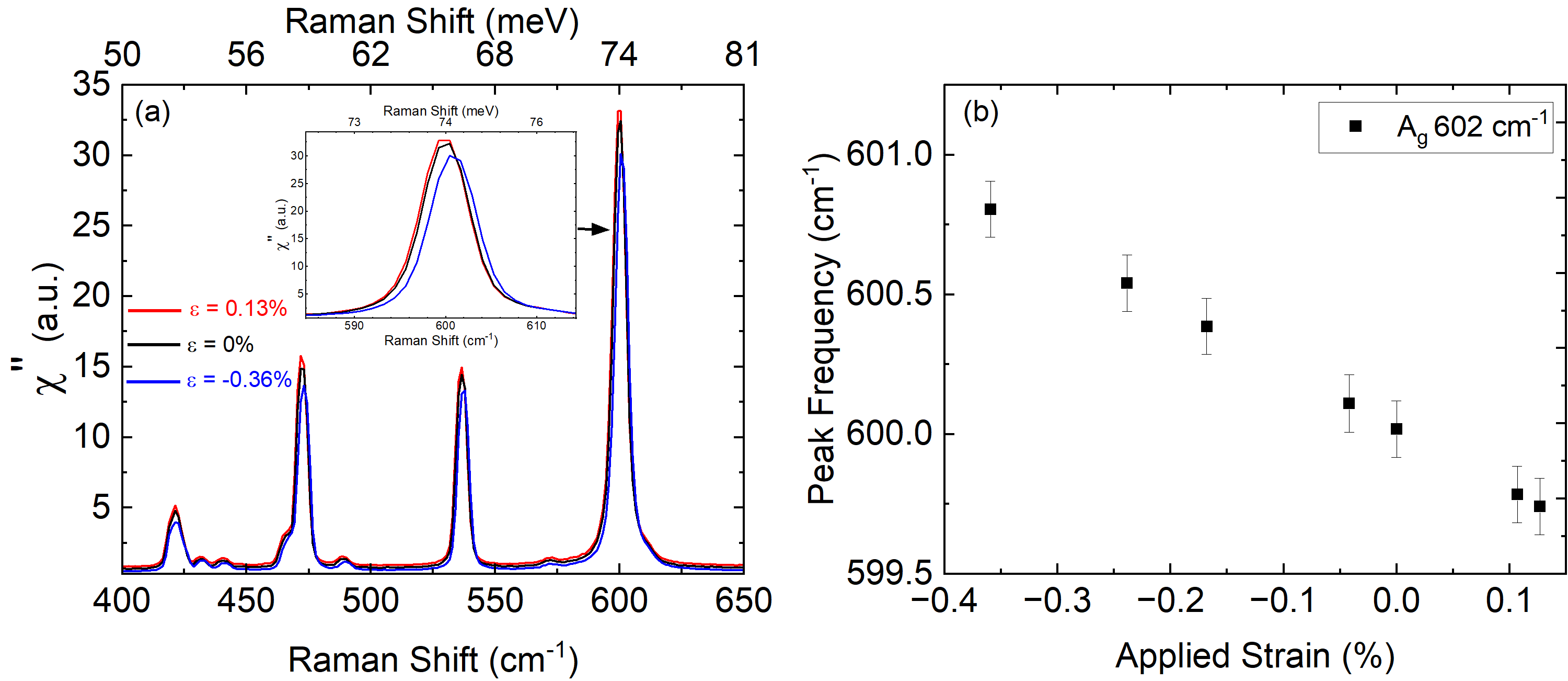} 
    \caption{(a) Unpolarized Raman spectra of $\alpha$-SrCr$_2$O$_4$ at 35 K in the strain cell at frequencies above 400 cm$^{-1}$ at maximum tensile pressure, zero pressure, and maximum compressive pressure. Inset: A$_g$ 602 cm$^{-1}$ phonon. (b) Peak Frequency of A$_g$ 602 cm$^{-1}$ phonon as a function of applied uniaxial pressure}
    \label{fig:strain_35K_high_freq} 
\end{figure}

In Fig. \ref{fig:strain_35K_high_freq} we present spectra of \sco ~obtained at T = 35 K under applied uniaxial pressure in the high frequency range {\it in-situ}. Maximum compression and tension spectra, showing the shifts of phonons under uniaxial pressure, are presented in Fig. \ref{fig:strain_35K_high_freq}(a) with negative signs denoting compressive pressure (all spectra are provided in the Supplemental Materials). The dependence of the phonon frequency on uniaxial pressure is reported in Fig. \ref{fig:strain_35K_high_freq}(b), where strain values are estimated from the calculated gauge factor above. Apart from a shift of frequency, an absence of other changes in the phonon spectrum such as phonon splitting or the appearance of new phonons indicates an absence of structural transitions under applied uniaxial pressure in the studied range of strain values. 

\section{\label{two-magnon}Two-magnon Scattering and Magneto-elastic Coupling as a Function of Temperature and Uniaxial Pressure}

\begin{figure}
    \centering
    \includegraphics[width=\linewidth]{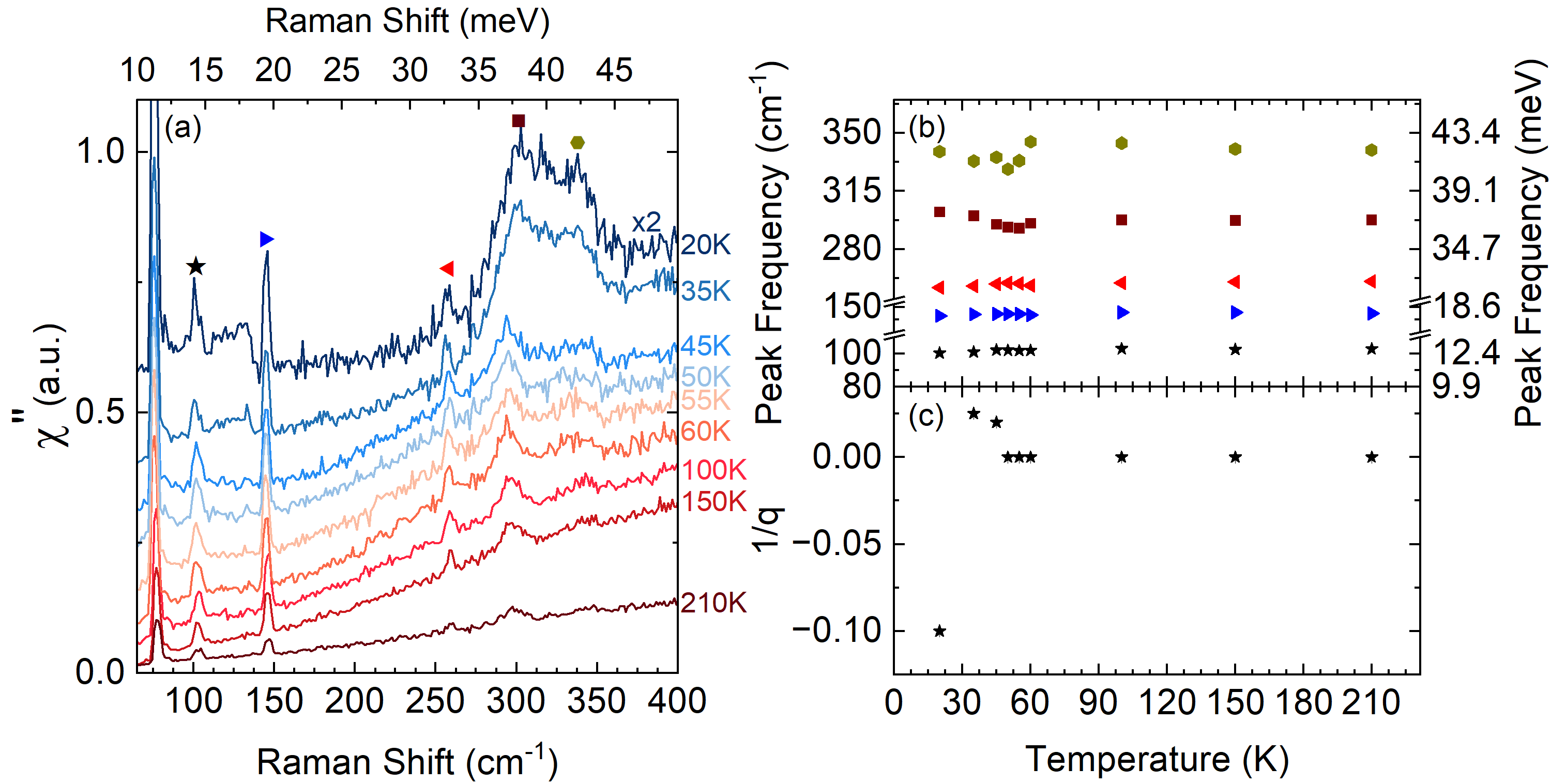} 
    \caption{(a) Unpolarized temperature dependent Raman spectra of $\alpha$-SrCr$_2$O$_4$ at frequencies below 400 cm$^{-1}$ in the strain cell. Spectra shifted along the y axis for clarity. (b) Frequencies of select oxygen-related phonons marked in panel (a). (c) Temperature dependence of the electron-phonon coupling strength, 1/q, of A$_g$ 102 cm$^{-1}$ phonon.}
    \label{fig:tdep_low_freq}
\end{figure}

In Fig. \ref{fig:tdep_low_freq} we present data obtained in the 65- 400~\cm ~frequency range from 210 K to 20 K for the sample cooled down in the strain cell. Below T$_N$ we observe a development of two broad spectral features at 125 and 325 cm$^{-1}$. The spectra overall are similar to the ambient pressure data despite the tensile stress of 0.41\% applied to the sample due to differential thermal expansion. These broad excitations observed below T$_N$ were attributed to two-magnon scattering in \cite{Valentine2015}, with the two features originating from the flat part of the magnon dispersion \cite{Brenig2008}, as is confirmed by our two-magnon density of states calculations presented in this manuscript in Section \ref{theory}. Above 43 K, the continuum is still present up to 3T$_N$, since Raman scattering Fleury-Loudon response is a probe of short-range magnetic correlations \cite{Loudon1968}; however, the magnetic continuum above T$_N$ does not show the well-defined maxima related to the flat part of the spin wave dispersion. 

Phonons in this frequency range visibly interact with the emerging two-magnon excitations as evidenced by their line shapes which become more asymmetric as the two-magnon scattering intensity increases. This asymmetry can be described by a phenomenological coupling parameter, $1/q$, when fit with a Fano line shape \cite{Fano1961,Tan2019,Ding2015}, where the Fano line shape is described with the following formula:

\begin{equation}
    F(\omega, \omega_0, \Gamma_F, q) = I_0\frac{[1+ 2(\omega-\omega_0)/q\Gamma_F ]^2}{1 + 4(\omega-\omega_0)^2/\Gamma_F^2}
\end{equation}

where $I_0$, $\omega_0$, and $\Gamma_F$ are the intensity, uncoupled phonon frequency, and broadening parameter, respectively. In Fig. \ref{fig:tdep_low_freq}(d), $1/q$ for the A$_g$ 102 cm$^{-1}$ phonon is plotted as a function of temperature. The coupling parameter, $1/q$, becomes negative on cooling down in the cell ($\epsilon$=0\%) below T$_N$ indicating the presence of a low frequency component part of the two-magnon continuum. 

\begin{figure*}[t]
    \centering
    \includegraphics[width=\linewidth]{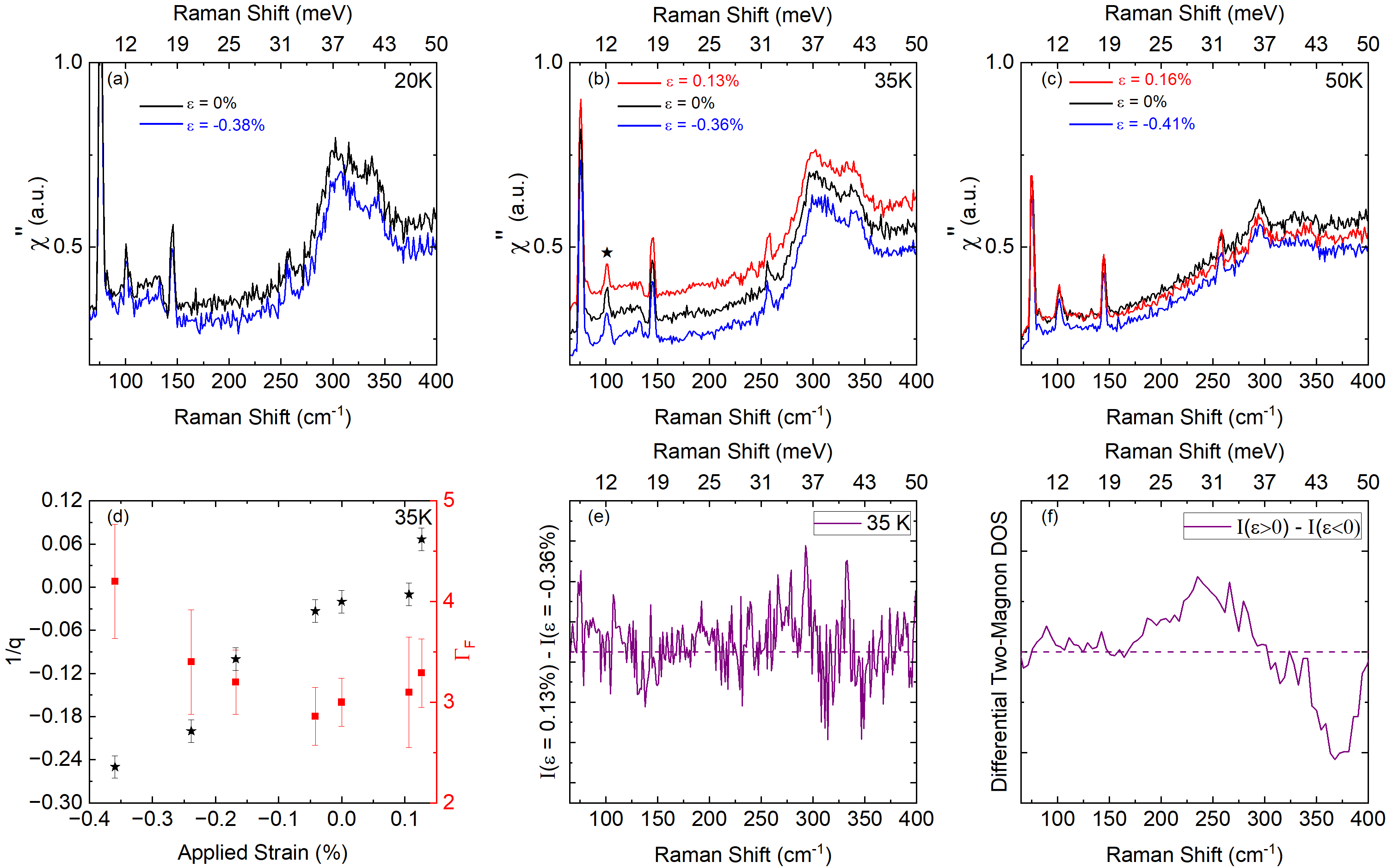} 
    \caption{(a) Unpolarized Raman spectra of $\alpha$-SrCr$_2$O$_4$ in the strain cell at frequencies below 400 cm$^{-1}$. Spectra taken at 20 K under applied uniaxial pressure. (b) Spectra taken at 35 K under applied uniaxial pressure in the same frequency range as (a). Spectra offset vertically for clarity. (c) Spectra taken at 50 K under applied uniaxial pressure in the same frequency range as (a). (d) Left axis: Electron-phonon coupling strength, 1/q, of A$_g$ 102 cm$^{-1}$ phonon at 35 K as a function of applied uniaxial pressure. Right axis: Broadening parameter $\Gamma_F$ of A$_g$ 102 cm$^{-1}$ phonon as a function of applied uniaxial pressure at 35 K. (e) Difference between the intensities at maximum tension and maximum compression applied to the sample, I($\varepsilon$ = 0.13\% )-I($\varepsilon$ = -0.36\%) at 35~K (b). (f) Differential two-magnon density of states calculation, $I(\varepsilon >$ 0) - $I(\varepsilon <$ 0).}
    \label{fig:strain_35K}
\end{figure*} 

Next, we follow the change in the two-magnon excitations of \sco ~under applied uniaxial pressure. At 50 K, above T$_N$, spectra have been obtained at the following values: -0.41\%, -0.2\%, -0.1\%, -0.04\%, 0\%, 0.08\%, and 0.16\%, where 0\% corresponds to the absolute strain of $\sim$0.41\% achieved upon cooling the sample in the strain cell due to differential thermal expansion. While the continuum of magnetic excitations is present in the spectra up to 3T$_N$ due to strong thermal fluctuations, the changes are below the sensitivity of our measurements. We observe almost no difference between the spectra measured at -0.41~\% and 0.16~\% of applied uniaxial pressure (see Fig.~ \ref{fig:strain_35K}(c)). 

At 35 K, spectra have been obtained at the following values: -0.36\%, -0.24\%, -0.17\%, -0.04\%, 0\%, 0.11\%, and 0.13\% (see Fig.\ref{fig:strain_35K}(b)) 
The largest changes as a function of uniaxial pressure are detected for the lower-frequency maximum of the two-magnon response, below 150 \cm.  We observe a tendency of the lower frequency part of the two-magnon continuum to shift to higher frequencies as uniaxial pressure decreases from 0.13\% applied tensile pressure to -0.36\% applied compressive pressure (see Fig. \ref{fig:strain_35K} (b)). The distinct Fano shape of the A$_g$ phonon at 102 ~\cm\ for $\epsilon <$ 0.13 \% suggests the presence of the underlying continuum. The change in the value and the sign of the coupling parameter 1/q as a function of pressure indicates the shift of the continuum in this frequency range (see Fig. \ref{fig:strain_35K} (d)) or an essential change in magneto-elastic coupling. 

The effect of uniaxial pressure on the maximum of the two-magnon excitation around 325~\cm can be deduced from observing the change in asymmetry of the A$_g$ 298 \cm\ and B$_{3g}$ 346 \cm\ phonons coupled to the magnetic continuum \cite{Valentine2015}. We observe the B$_{3g}$ 346 \cm\ phonon narrow and lose spectral weight while the A$_g$ 298 \cm\ phonon broadens and increases in spectral weight as applied uniaxial pressure decreases from 0.13\% to -0.36\% suggesting a shift to lower frequencies as applied uniaxial pressure decreases (see Fig. \ref{fig:strain_35K}(e)). 

At 20 K, we observe the same effect discernible at $35~K$. In the higher-frequency maximum part of the two-magnon excitation spectrum around 325 \cm, we interpret the line shape changes as the result of the shift of the spectral weight of the two-magnon excitations to higher frequencies as applied uniaxial increases from $\epsilon =$ -0.36 \%  to $\epsilon =$ 0.13 \% (see Fig. \ref{fig:strain_35K}(a) and \ref{fig:strain_35K}(e)). 

\section{\label{theory}Spectrum of Two-magnon Excitations: Spin wave calculations}

\begin{figure*}[t]
    \centering
    \includegraphics[width=\linewidth]{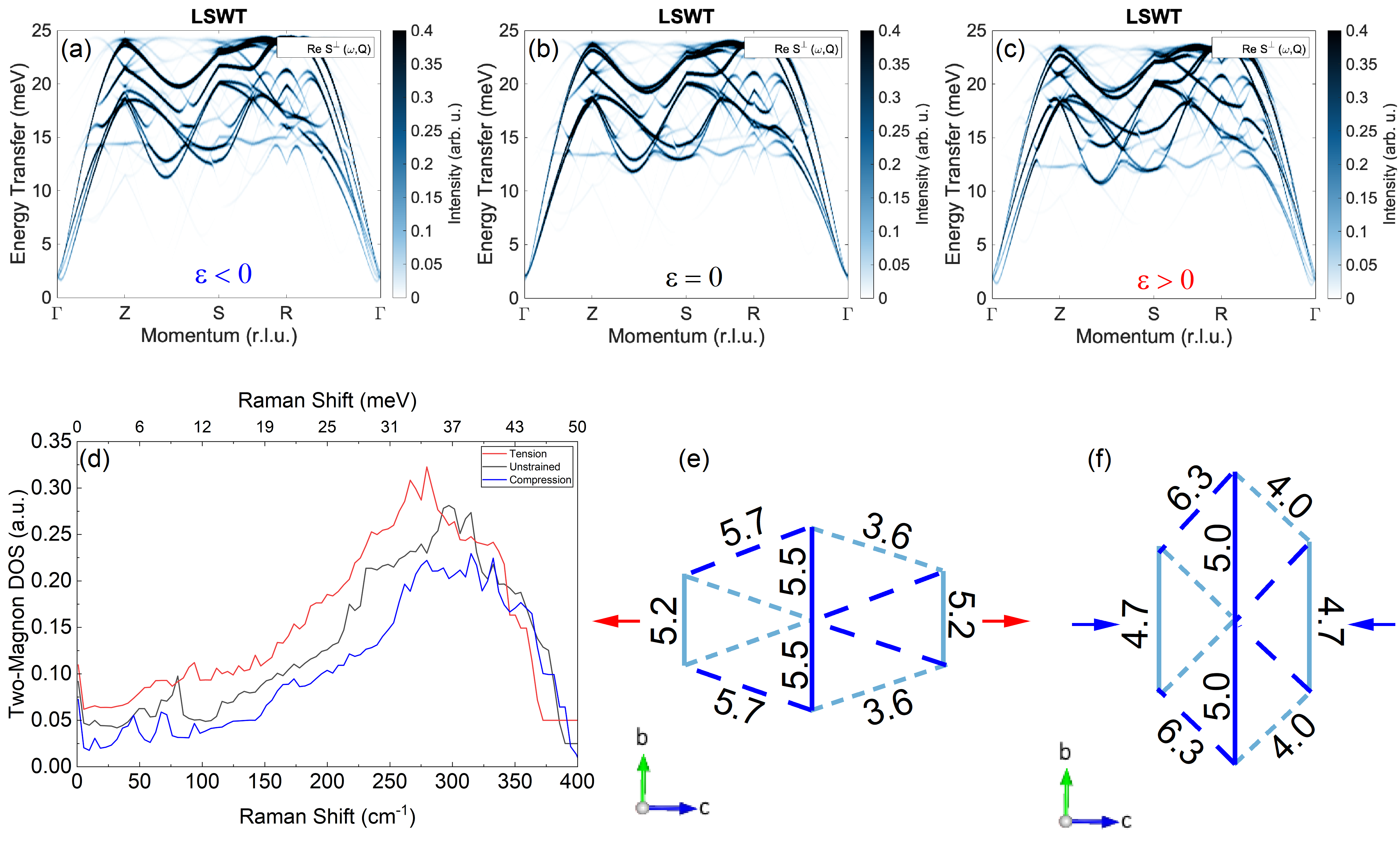}
    \caption{(a)-(c) Linear Spin Wave Theory (LSWT) calculations for $\alpha$-SrCr$_2$O$_4$ along high symmetry points in orthorhombic Brillouin Zone for applied compressive pressure, zero pressure, and applied tensile pressure, respectively. (d) Resulting two-magnon density of states. Calculations offset vertically for clarity. (e) Simplified schematic view of exchange interactions on triangular lattice of Cr$^{3+}$ under applied tensile uniaxial pressure along the c axis. (f) Same as (e) under applied compressive uniaxial pressure.}
    \label{fig:spinwave_strain}
\end{figure*}

To connect our experimental observations with theoretical expectations, we use spin wave calculations to obtain the magnon dispersion of $\alpha$-SrCr$_2$O$_4$ as a function of applied uniaxial pressure and subsequently model magnetic Raman scattering spectra. Albeit \sco ~has almost structurally perfect triangular layers, in-equivalent Cr$^{3+}$ ions lead to four distinct nearest-neighbor interactions, J$_{ch1}$, J$_{ch2}$, J$_{zz1}$, and J$_{zz2}$, as was shown by DFT calculations \cite{Valentine2015}, which are marked in Fig. \ref{fig:structure}(c). Both direct cation-cation magnetic exchange interactions and superexchange interactions through O$^{2-}$ contribute to these exchange interactions \cite{Valentine2015}; however, the corresponding angle in the Cr-O-Cr superexchange pathway is $\sim$90$^\circ$ resulting in a fairly weak superexchange interaction \cite{Goodenough1960}. Parameters for nearest-neighbor interactions were extracted from experimental neutron scattering results in \cite{Damay2017}. While the easy-plane anisotropy has not been estimated for \sco ~in literature, given that the magnetic ground state is fairly close to the isotropic TLHAF, we can estimate a relatively small anisotropy. By varying the four nearest-neighbor interactions by $\sim$5\%, which would lead to more isotropic triangular lattice ($\epsilon <$ 0) or more anisotropic ($\epsilon >$ 0), we were able to demonstrate how uniaxial pressure applied along the $c$ in-plane axis affects the two-magnon density of states in $\alpha$-SrCr$_2$O$_4$ (see Fig. \ref{fig:spinwave_strain}). 

Values for all exchange interactions and the magnitude of anisotropy used for simulated spinwave calculations are seen in Table \ref{table:J}. Using the following magnetic Heisenberg Hamiltonian defined by Damay \textit{et al.} \cite{Damay2017} and Williams \textit{et al.} \cite{Williams2011}, the spinwave dispersion and two-magnon density of states of $\alpha$-SrCr$_2$O$_4$ were simulated using SpinW \cite{Lake2015}: 

\begin{equation}
    \mathcal{H} = \mathcal{H}_{nn} + \mathcal{H}_{aniso},
\end{equation}

where

\begin{multline}
\mathcal{H}_{nn} = \sum_{i,j}J_{ch1}\boldsymbol{S_{ij}^{1a}}\cdot(\boldsymbol{S_{ij}^{1b}} + \boldsymbol{S_{(i-1)j}^{1b}})+ \\J_{ch2}\boldsymbol{S_{ij}^{2a}}\cdot(\boldsymbol{S_{ij}^{2b}} + \boldsymbol{S_{(i-1)j}^{2b}})+ J_{zz1}\boldsymbol{S_{ij}^{1a}}\cdot(\boldsymbol{S_{i(j-1)}^{2a}} + \boldsymbol{S_{i(j-1)}^{2b}}) + \\J_{zz1}\boldsymbol{S_{ij}^{1b}}\cdot(\boldsymbol{S_{(i+1)j}^{2a}} + \boldsymbol{S_{ij}^{2b}}) + J_{zz2}\boldsymbol{S_{ij}^{1a}}\cdot(\boldsymbol{S_{ij}^{2a}} + \boldsymbol{S_{ij}^{2b}}) + \\J_{zz2}\boldsymbol{S_{ij}^{1b}}\cdot(\boldsymbol{S_{(i+1)(j-1)}^{2a}} + \boldsymbol{S_{i(j-1)}^{2b}})
\end{multline}

and

\begin{equation}
\begin{split}
\mathcal{H}_{aniso} = \sum_{i}D\boldsymbol{(S_{i}n)^2}.
\end{split}
\end{equation}

$\boldsymbol{S_{ij}^\alpha}$ is the spin of the Cr$^{3+}$ site ($\alpha \in \{1a,1b,2a,2b\}$), and i and j are indices of the unit cell along the \textit{b} and \textit{c} axes, respectively (see Fig. \ref{fig:structure}(b)). $D$ denotes an easy-plane anisotropy, and $\boldsymbol{n}$ is a vector perpendicular to this plane as helical magnetic order develops with spins confined in the \textit{ac} plane. 

While the resulting spinwaves seen in Fig. \ref{fig:spinwave_strain} are very similar to the isotropic TLHAF \cite{Brenig2008,Zhitomirsky2015}, lattice distortions due to in-equivalent Cr$^{3+}$ sites lift classical degeneracy leading to a softening of the magnon branch under ambient conditions \cite{Guillou1989,Damay2017}. We also observe additional magnon branches in the dispersion due to crystallographic twinning in the \textit{bc} plane \cite{Damay2017}.  

Two magnon density of states, which corresponds to the magnetic Raman scattering signal is calculated using the obtained spin wave dispersion (see Fig. \ref{fig:spinwave_strain}(b)). Details of the two-magnon density of states calculation are discussed in the Supplemental Materials. Based on the calculation, we expect one broad feature around $\sim$ 37 meV (360 cm$^{-1}$) and another smaller broad feature around $\sim$ 9 meV (70 cm$^{-1}$) to emerge below T$_N$, which agrees with the published \cite{Valentine2015} spectra of \sco\ and those obtained in this work. 
 As a general trend we observe that compression ($\epsilon <$ 0) reduces the anisotropy, and pushes high-frequency two-magnon peak to higher frequencies, and lower one to the lower frequencies, while tension ($\epsilon >$ 0) increases the anisotropy, resulting in the opposite behavior of the two peaks. 

\begin{table}
    \centering
    \begin{tabular}{|c|c|c|c|c|}
        \hline
                   & Exchange path & Compression & No Pressure & Tension\\
                   &   & J (meV) & J (meV) & J (meV)\\
        \hline
        J$_{ch1}$  &  Cr1 - Cr1  &  4.95 & 5.21 & 5.47\\
        J$_{ch2}$  &  Cr2 - Cr2  &  4.66 & 4.91 & 5.15\\
        J$_{zz1}$  &  Cr1 - Cr2  &  4.01 & 3.82 & 3.63\\
        J$_{zz2}$  &  Cr1 - Cr2  &  6.32 & 6.02 & 5.72\\
        D$_{aniso}$ &  NA        &  0.035 & 0.035 & 0.035\\ 
        \hline
    \end{tabular}
    \caption{Input parameters for calculated two-magnon density of states presented in Fig. \ref{fig:spinwave_strain} for applied compression, no applied pressure, and applied tension, respectively.}
    \label{table:J}
\end{table}

\section{\label{discussion}Discussion}

The main objective of this work was to explore the magnetic ground state of a triangular lattice antiferromagnet \sco\ (T$_N$= 43~K) by tuning anisotropy away from $J_2/J_1 \sim$ 0.82 through in-situ application of uniaxial pressure. 
We used the frequencies of Raman active oxygen phonons to determine strain transferred to the sample, while simultaneously measuring magnetic Raman response. Due to the difference in thermal expansion coefficients of the sample and Ti plates of the strain cell, 0.41\% of thermal strain at low temperatures was applied to the sample with zero voltage applied to the piezos of the strain cell. At low temperatures, uniaxial pressure was applied {\it in-situ} in the range of -0.41\% to 0.16\%. While there was an interest to study the response of \sco ~in a broader range of strain values, the experimental parameters were limited by the elastic limit of the material, size of the sample, and operational limit of the strain cell. 

The spectrum of two-magnon excitations  present below T$_N$=43~K  is in the overall agreement with the previous results on magnetic Raman scattering of \sco\ at  ambient pressure ~\cite{Valentine2015}. Observed energies of two-magnon excitations  are in agreement to our spin wave calculations based on the parameters obtained by the fit of neutron scattering data \cite{Damay2017}. 

Changes in magnetism on the application of uniaxial pressure resulted in  subtle but distinct changes of two-magnon excitations in the spectra  below T$_N$ at 35 and 20~K, with larger amplitude changes detected at the temperature closer to the transition. These changes are in agreement with the expectations provided by our spin wave calculations for the change of two-magnon excitations under applied uniaxial pressure (see Fig. \ref{fig:spinwave_strain}), where under compression along the in-plane $\textit{c}$ axis we expect a more isotropic triangular lattice resulting in the two maxima of the two-magnon excitation continuum moving apart from each other, with the lowest shifting to lower frequency.  
At 35 K, the largest change is observed for the lower frequency maximum of the two-magnon response, which shifts from $\sim$100 \cm\ under tension ($\varepsilon$=0.13\%) to 130 \cm\ under compression ($\varepsilon$=-0.36\%) within the accessible strain range. We also observe distinct changes in the line shapes of magneto-elastically coupled phonons under applied uniaxial pressure.  

We speculate how uniaxial pressure would influence the different exchange paths and mechanisms. When tensile uniaxial pressure is applied along the \textit{c} axis, the direct Cr-Cr exchange along the \textit{b} axis increases while the Cr-O-Cr superexchange angle $\phi$ decreases resulting in a decrease of the antiferromagnetic superexchange interaction (see Fig. \ref{fig:spinwave_strain}(e)). While the superexchange interaction decreases, the dominant magnetic interactions in this material interact by direct exchange as the superexchange interaction is fairly weak. $J_{ch2}/J_{zz2}$ also increases from $\approx$ 0.82 to 1.03 under applied uniaxial tension as magnetic interactions are tuned by $\sim$ 5\% indicating an increase in frustration. The resulting softening of the two-magnon excitation seen in Fig. \ref{fig:spinwave_strain} thus agrees with previous theoretical predictions of anisotropic TLHAFs. Literature predicts frustration-induced softening of the magnon dispersion as $J_{2}/J_{1}$ increases which would present as a spectral downshift of the magnetic Raman feature in the spectra \cite{Guillou1989,Gingras2007}. 

When compressive uniaxial pressure is applied along the \textit{c} axis, the direct Cr-Cr exchange along the \textit{b} axis decreases while the superexchange interaction increases (see Fig. \ref{fig:spinwave_strain}(f)). In Fig. \ref{fig:spinwave_strain}(d), the high energy feature of the two magnon excitation shifts from low to high frequencies as uniaxial pressure decreases agreeing with the predicted response. Here $J_{ch2}/J_{zz2}$ decreases from $\approx$ 0.82 to 0.67 under applied uniaxial compression indicating a decrease in frustration. This would imply a spectral upshift and a hardening of the magnetic Raman feature in the spectra as frustration $J_{2}/J_{1}$ decreases.  

Due to the Poisson ratio, assumed to be 0.2 \cite{Son2016}, the effect under applied tensile pressure along the \textit{c} would be similar to when compressive pressure is applied along the \textit{b} axis.
Given that our experimental two-magnon response under uniaxial pressure is opposite to theoretical calculations, we deduce that uniaxial pressure is applied along the in-plane $\textit{b}$ axis as opposed to the in-plane $\textit{c}$ axis.  






\section{\label{conclusion} Conclusion and Outlook}

We present an experimental study of {\it in-situ} tuning of the lattice and magnetic excitations in the quasi-2D anisotropic triangular lattice antiferromagnet $\alpha$-SrCr$_2$O$_4$ under uniaxial pressure. Within the tuning range of about 0.5\% of the unit cell allowed by the technique and mechanical properties of the material, we observe a subtle shift of two-magnon excitations in agreement with spin wave calculations. More pronounced changes of magneto-elastic coupling on tuning uniaxial pressure at temperatures just below T$_N$ can be understood in terms of tuning lattice symmetry parameters related to ferroelectricity.  

While the magnetic interactions are tuned within the helical magnetic phase of $\alpha$-SrCr$_2$O$_4$, the presented results provide insight into how the application of uniaxial pressure in-plane vs out-of-plane may affect the magnetic structure. Given that the helical magnetic structure in $\alpha$-SrCr$_2$O$_4$ has spins confined to the \textit{ac} plane, minimal uniaxial pressure applied in-plane would only distort the spin helix structure in contrast to uniaxial pressure applied out-of-plane, which could cause a destabilization of the long-range magnetic order \cite{Wu2012} and lead to possible magnon decay \cite{Zhitomirsky2006} or a different magnetic order. 

Additionally, the presented results provide insight into how other materials in the $\alpha$-MCr$_2$O$_4$ (M = Ca, Sr, Ba) family might behave under uniaxial pressure. $\alpha$-SrCr$_2$O$_4$ is closer to the ideal hexagonal lattice compared to $\alpha$-CaCr$_2$O$_4$ \cite{Wu2012}. Given that the Ca compound is fairly close to a phase boundary, tuning the material out of the helical magnetic phase might be feasible, and the resulting frustration-induced softening of the two-magnon excitation could appear more pronounced.  

\section{Acknowledgments}

This research has been funded by the United States Department of Defense (DoD) under Grant No. W911NF2120213.

\bibliography{references}

\end{document}


\title{Supplemental Material for: Tuning J$_1$-J$_2$ in Quasi-2D Triangular Lattice Antiferromagnet $\alpha$-SrCr$_2$O$_4$ via Uniaxial Pressure}
\author[1]{Jazzmin Victorin} 
\author[1]{Shreenanda Ghosh}
\author[1]{Seyed Koohpayeh}
\author[1]{Chris Lygouras}
\author[1]{Natalia Drichko}

\affil[1]{Department of Physics and Astronomy, Johns Hopkins University, Baltimore, Maryland 21218, USA}

\maketitle

\section{Raman Spectroscopy} 

\subsection{Fitting Raman Spectra}

Raman spectra were fit with a sum of Lorentzian line shapes given by Eq. \ref{eq:lorentz} as well as Gaussian-Lorentzian (pseudo-Voigt) lineshapes given by Eq. \ref{eq:voigt}. 

\begin{equation}
   L(x) = \frac{a}{1+(\frac{x-\omega}{\Gamma/2})^{2}}
    \label{eq:lorentz}
\end{equation}

\begin{equation}
   V(x) = \eta\frac{1}{2}G(x) + (1-\eta)L(x) = \eta e^{-\ln2(\frac{x-\omega}{\Gamma/2})^2} + (1-\eta)\frac{a}{1+(\frac{x-\omega}{\Gamma/2})^{2}}
    \label{eq:voigt}
\end{equation}

Here a is the height of the peak, $\omega$ is the peak frequency, $\Gamma$ is the full width at half max, and $\eta$ is the mixing parameter.

\begin{figure}
    \centering
    \includegraphics[width=0.5\linewidth]{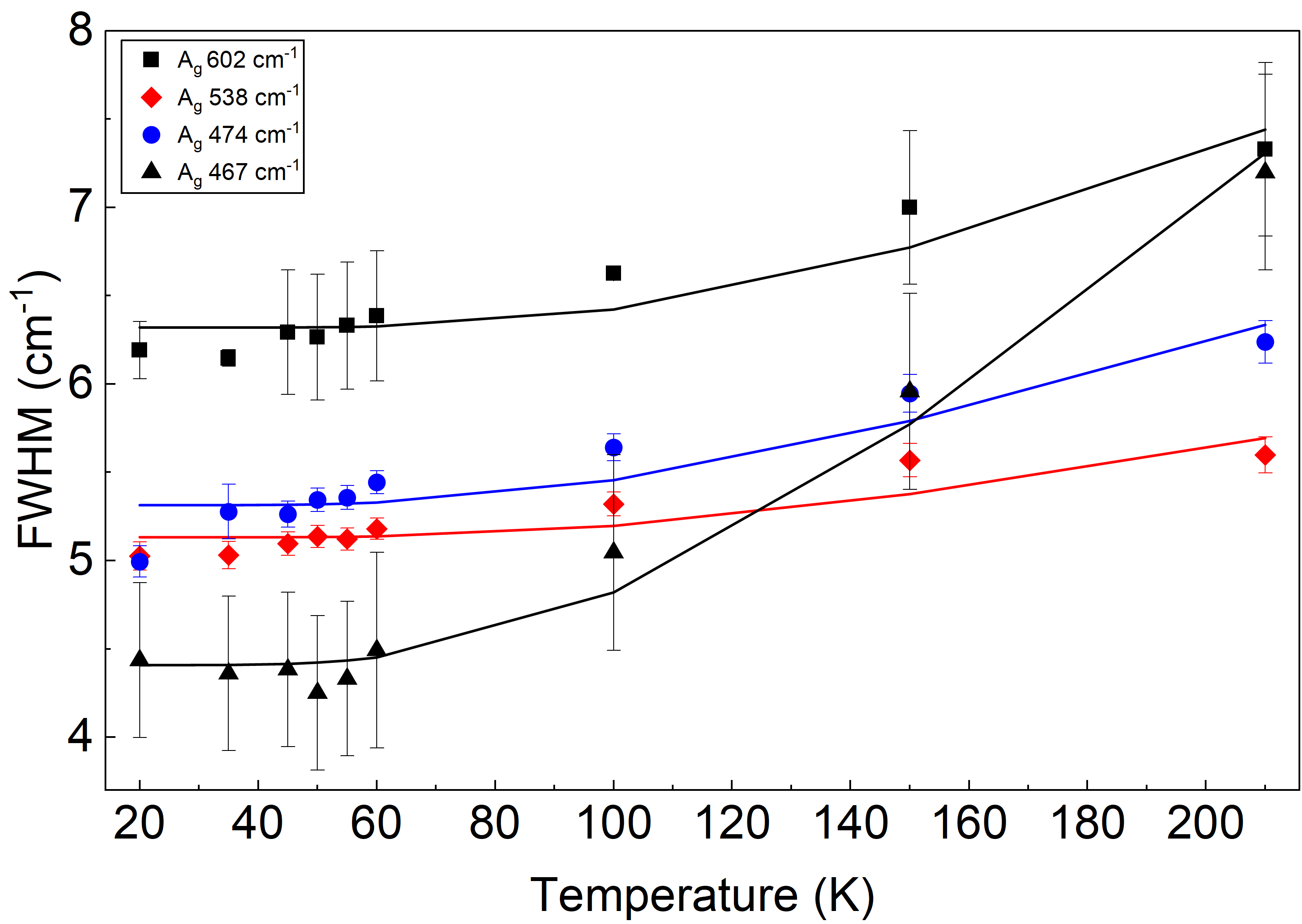} 
    \caption{Phonon widths of select oxygen-related phonons as a function of temperature. Solid lines are fits to Klemens model (Eq. \ref{eq:klemens}).}
    \label{fig:widths}
\end{figure}

In contrast to the difference in the temperature dependence of the phonon positions, the temperature dependence of the width of the phonons is similar in both ambient and strain experiments and follows the typical phonon-phonon scattering behavior as a function of temperature. This can be seen in Fig. \ref{fig:widths} where solid lines are fits to the following model: 

\begin{equation}
   \Gamma(T) = \Gamma_0 + A(1 + 2n(\omega,T))
    \label{eq:klemens}
\end{equation}

where $\Gamma_0$ is a temperature independent term, and A is a fitting parameter \cite{Klemens1966}. 

\subsection{Gr\"uneisen Parameter for Select High Frequency Phonons}
 
\begin{table*}[h]
    \centering
    \begin{tabular}{|c|c|c|}
        \hline
                             &  Gr\"uneisen Parameter ($\gamma$) & Gauge Factor (\cm/\%)  \\
        \hline
        A$_g$ 602 cm$^{-1}$  &  0.45 $\pm$ 0.18 & 2.19 \\
        B$_{3g}$ 575 cm$^{-1}$  &  1.49 $\pm$ 0.42 & 6.95 \\
        A$_{3g}$ 538 cm$^{-1}$  &  0.55 $\pm$ 0.17 &  2.36 \\
        B$_{3g}$ 490 cm$^{-1}$  &  2.60 $\pm$ 0.82 & 10.12 \\
        A$_g$ 474 cm$^{-1}$     & 2.03 $\pm$ 0.78 &  7.85 \\
        A$_g$ 467 cm$^{-1}$     & 6.81 $\pm$ 2.08 & 26.79 \\
        
        \hline
    \end{tabular}
    \caption{Extracted Gr\"uneisen parameters for select high frequency phonons are listed along with their resulting gauge factors.}
    \label{table:gauge}
\end{table*}



        

\section{Two-Magnon Density of States Calculations}

The two-magnon density of states (DOS) calculations require details of the spin wave modes. These can be estimated from linear spin wave theory (LSWT). We consider the Fourier transform of the exchange interactions, where $J_A(q)$ are for interactions between atoms in the same plane, and $J_{AB}(q)$ are the interactions between adjacent planes. The values of the exchange interactions for arbitrary-distance interaction scale can be estimated from the inelastic neutron scattering spectrum seen in \cite{Damay2017}. For two-magnon scattering, we consider the tensor

\begin{equation}
    \tilde{R}(\vec{k}) = S^\dagger_{\vec{k}}\tilde{R_0}(\vec{k})S_{\vec{k}}, 
    \label{eq:2Mtensor}
\end{equation}

where $\tilde{R_0}(\vec{k})$ is a 4 x 4 tensor related to the Fourier transform $\sum_j (\vec{E}_{in}\cdot\hat{\delta}_{ij})(\vec{E}_{out}\cdot\hat{\delta}_{ij})e^{i\vec{k}\cdot \vec{\delta}_{ij}}$.
The scattering intensity is proportional to the imaginary part of the Green’s function,

\begin{equation}
    I(\omega) \propto -Im \sum_{\nu,\mu=1,2} \sum_{\vec{k}}\frac{\tilde{R}_{\mu+2,\nu}(\vec{k})\tilde{R}_{\mu,\nu+2}(\vec{k})}{\omega-(\omega_\mu+\omega_\nu)+i\varepsilon}, 
    \label{eq:Greenfunc}
\end{equation}

where the sum $\nu,\mu$ is over the two-magnon branches, and the sum over the wave vector is taken over the first Brillouin Zone (BZ).

\printbibliography